\documentstyle[12pt,aasms] {article}
\begin{document}

\def\vecv{{\bf v}}

\def\vecB{{\bf B}}
\def\vecE{{\bf E}}
\def\vecJ{{\bf J}}

\def\vecvi{{\bf v}_i}
\def\vecvn{{\bf v}_n}
\def\vecvD{{\bf v}_D}

\def\Fin{F_{in}}
\def\Hin{H_{in}}
\def\nuin{\nu_{in}}
\def\nuni{\nu_{ni}}
\def\rhoi{\rho_i}
\def\rhon{\rho_n}
\def\RAD{R_{AD}}

\title{Ambipolar Drift Heating in Turbulent Molecular Clouds}
\author{Paolo Padoan}
\affil{Harvard University Department of Astronomy, Cambridge, MA 02138,
padoan@cfa.harvard.edu}
\author{ Ellen Zweibel} 
\affil{JILA, University of Colorado at Boulder, Boulder CO 80309-440,
zweibel@solarz.colorado.edu}
\author{{\AA}ke Nordlund}
\affil{Astronomical Observatory and Theoretical Astrophysics Center, Juliane
Maries Vej 30, DK-2100 Copenhagen, Denmark, aake@astro.ku.dk}

\begin{abstract}

Although thermal pressure is unimportant dynamically in most molecular gas,
the temperature is an important diagnostic of dynamical processes and
physical conditions. This is the first of two papers on thermal equilibrium
in molecular clouds. We present calculations of frictional heating by 
ion-neutral (or ambipolar) drift in three--dimensional simulations of turbulent,
magnetized molecular clouds. 

We show that ambipolar drift heating is a strong function of position 
in a turbulent cloud, and its average value can be significantly
larger than the average cosmic ray heating rate. 
The volume averaged heating rate per unit volume due to ambipolar drift, $H_{AD}
=\vert{\bf J\times B}\vert^2/\rho_i\nu_{in}
\sim B^4/(16\pi^2 L_B^2\rho_i\nu_{in})$, is found to depend on
the rms Alfv\'{e}nic Mach number, 
${\cal M}_A$, and on the average field strength, as
$H_{AD}\propto {\cal M}_A^2\langle |B|\rangle ^4$. This implies that 
the typical
scale of variation of the magnetic field, $L_B$, is inversely
proportional to ${\cal M}_A$, which we also demonstrate.

\end{abstract}

\section{Introduction}

Observations reveal that roughly 20\% - 40\% of the interstellar gas in the
Galactic disk is organized into Giant Molecular Clouds (GMCs). These clouds
appear to be self gravitating and roughly in virial equilibrium. The density
distribution in the clouds 
is highly inhomogeneous, and the velocity field is turbulent
and highly supersonic. It is thought that both magnetic and turbulent
kinetic pressure contribute to cloud support, but the relative degree of
magnetic support is uncertain because of the observational difficulty of
measuring the magnetic field strength and topology.

Several lines of evidence suggest that GMCs survive for a few tens of 
millions of years, but
both analytical estimates, some dating back more than two decades (Goldreich
\& Kwan 1974, Field 1978, Zweibel \& Josafatsson 1983, Elmegreen 1985),
and numerical simulations (Scalo \& Pumphrey 1982, Stone, Ostriker, \& Gammie
1998, MacLow 1999, Padoan \& Nordlund 1999) suggest that molecular cloud turbulence
dissipates in much less than the cloud lifetime.
Star formation is observed to take place in nearly all GMCs, and it appears
that the lifetimes of GMCs are determined by the rate at which they are
destroyed by energy and momentum input from the massive stars which form
within them. It also appears likely that energy input from stars, and possibly
from other sources, drives turbulence in GMCs as well.

Ambipolar drift, or ion-neutral friction, has long been thought to be an 
important energy dissipation mechanism in molecular clouds, and therefore a
significant heating mechanism for molecular cloud gas (Scalo 1977, Goldsmith
\& Langer 1978, Zweibel \& Josafatsson 1983, Elmegreen 1985). In fact, as first
suggested by Scalo (1977), the observed low temperatures of molecular clouds
places upper limits on the rate of energy dissipation by ambipolar drift, and
thus, with some additional assumptions, on the magnetic field strength.

Nevertheless, it has been difficult to assess the rate of energy dissipation
by ambipolar drift. The frictional heating rate $H_{AD}$ depends on the 
local Lorentz force in the medium, which is almost impossible to measure.
Simple scaling arguments show that $H_{AD}$ is proportional to $B^4 n^{-1}
n_i^{-1}
L_B^{-2}$, where $B$, $n$, $n_i$, and $L_B$ are the magnetic field strength, 
neutral
density, ion density,
 and magnetic length scale, respectively. The extreme sensitivity of
this formula to magnetic field strength, only the line of sight average
value of which can be directly measured, and to
the essentially unobservable $L_B$, 
has made it difficult to estimate $H_{AD}$ with confidence even to
order of magnitude.

In this paper, we use numerical simulations of magnetized turbulence
to study heating by 
ambipolar drift in molecular clouds.
We show that the ambipolar heating rate per
unit volume, $H_{AD}$, depends on field strength as $B^4$, for constant
rms Mach number of the flow, and on the Alfv\'{e}nic Mach number
as ${\cal M}_A^2$. This implies that the magnetic length scale, $L_B$, depends
on the Alfv\'{e}nic Mach number as ${\cal M}_A^{-1}$,
which we demonstrate. We show that the numerical
value of the heating rate, computed by solving the three--dimensional
compressible magneto--hydrodynamic (MHD) equations, tends to converge
with increasing numerical resolution, and therefore we can fully quantify
the value of $H_{AD}$, to within an uncertainty of less than a factor of two,
in the numerical models.

These empirical formulae make it much easier to estimate ambipolar drift
heating in terms of observable properties of clouds. The average heating rate
depends on $B^2$, rather than $B^4$ as in the traditional expression, on
the line width or rms velocity, and on the neutral and ion densities. The
magnetic length scale is eliminated through its dependence on ${\cal M}_A$.

We find that ambipolar drift is probably a stronger heating mechanism than
cosmic ray heating, and that it is a strong damping mechanism for molecular
cloud turbulence, leading to significant decay within one dynamical
crossing time. 

In \S 2 of the paper we briefly summarize the physics of ambipolar drift
heating, and in \S 3 we describe the simulations. In \S 4 we present 
results of the computation of ambipolar drift heating rate, and in \S 5 
we discuss some implications of the results and summarize the conclusions
of this work.

\section{Dynamics and Dissipation of Weakly Ionized Gas}

Molecular cloud gas includes neutral atoms and molecules, atomic and molecular
ions, electrons, and dust grains, which may also be electrically charged. 
At the gas densities considered in this paper 
($0.1$~cm$^{-3}<n<10^5$~cm$^{-3}$, and $\langle n \rangle=320$~cm$^{-3}$), 
electrons are the primary current carriers, the
number density of ions much exceeds the number density of grains,
 and Ohmic dissipation is negligible (Nakano \& Umebayashi
1986). Moreover, significant charge separation cannot be sustained on the
time scales and length scales of interest, so the electrons and ions move
together. The electric field is then given by

\begin{eqnarray}
\label{E}
\vecE=-{{\vecvi}\over {c}}\times\vecB,
\end{eqnarray}
where $\vecvi$ is the velocity of the ion, or plasma component.

In principle, $\vecvi$ and the neutral velocity $\vecvn$ should be determined by
solving separate fluid equations for these species (Draine 1986), including
their coupling by collisional
processes. In typical molecular cloud environments this
leads to a great disparity of length scales and time scales, because ions collide
with neutrals at a very high frequency compared to other rates in the problem.
Past computations of two fluid systems (Toth 1994, Brandenburg \& Zweibel 1995,
Hawley \& Stone 1998) have avoided this problem by assuming ion to neutral
density ratios that are far higher than the 
values found in typical molecular clouds. An
alternative,  which we pursue here, is to consider only the
length scales and time scales over which the ions and neutrals are well coupled
(Shu 1983).

We introduce the drift velocity $\vecvD$

\begin{eqnarray}
\label{vD}
\vecvD\equiv\vecvi-\vecvn
\end{eqnarray}
and evaluate $\vecvD$ by assuming that the dominant forces on the charged
component are the Lorentz force and the frictional force arising from 
collisions with neutrals. We model the latter in the standard way as

\begin{eqnarray}
\label{Fin}
\Fin=-\rhoi\nuin\vecvD,
\end{eqnarray}
where $\rhoi\nuin=\rhon\nuni=\rhoi\rhon\langle\sigma v\rangle/(m_i+m_n)$. In
numerical work, we will take the collision rate coefficient 
$\langle\sigma v\rangle = 2\times 10^{-9}$
cm$^3$/s (Draine, Roberge, \& Dalgarno 1983) and assume the ions are HCO$^+$
(cf. de Jong, Dalgarno, \& Boland 1980)
and the neutrals are H$_2$ and He, with $n(He)/n(H_2)=2/9$. 
On time scales longer than the neutral-ion
collision time $1/\nuni$, the Lorentz and drag forces must balance. This leads
to an expression for the drift velocity

\begin{eqnarray}
\label{vDs}
\vecvD={{\vecJ\times\vecB}\over {c\rhoi\nuin}}.
\end{eqnarray}

Making the further assumption $\rho_i/\rho\ll 1$, so that $\vecvn$ is the same
as the center of mass velocity $\vecv$,
the magnetic induction equation is

\begin{eqnarray}
\label{ind}
{{\partial\vecB}\over {\partial t}} =\nabla\times (\vecvi\times\vecB)
= \nabla\times (\vecv\times\vecB) +
\nabla\times{{(\vecJ\times\vecB)\times\vecB}\over {c\rhoi\nuin}}.
\end{eqnarray}

The frictional heating rate $H_{AD}$ is

\begin{eqnarray}
\label{Hin}
H_{AD} = \rhoi\nuin\mid\vecvD\mid^2.
\end{eqnarray}

Equations (\ref{vDs}), (\ref{ind}), and (\ref{Hin}) 
constitute the strong coupling approximation
to ambipolar drift. They are much easier to solve, and
at less computational expense, than the full two fluid equations that hold for
even a single neutral species and a single charged fluid. Nevertheless, we
must consider where these equations break down.
The validity of the strong coupling approximation, even at low drift 
velocities, can be assessed by evaluating the ambipolar Reynolds number $\RAD$
(Zweibel \& Brandenburg 1997)

\begin{eqnarray}
\label{RAD}
\RAD\equiv{{L_Bv\nuni}\over {v_A^2}}\sim{{v}\over {v_D}},
\end{eqnarray}
where $L_B$ is the length scale on which the magnetic field varies,
and in the second relation we have used eqn. (4). The strong
coupling approximation should be valid for $\RAD \ga 1$. As examples, this
criterion gives the wavenumber for critical damping of small amplitude
MHD waves of wavelength $\lambda$ if we
replace $v$ by $v_A$ and $L_B$ by $\lambda/\pi$ in eqn. (\ref{RAD}) (McKee et 
al. 1993). And, the 
criterion for significant ion-neutral separation in shocks (Mullan 1971,
Draine 1980) is equivalent to the condition $\RAD < 1$ evaluated upstream.

We show in \S 4 that $L_B\propto{\cal M}_A^{-1}$, where ${\cal M}_A$ is an
appropriately averaged Alfven Mach number defined in eqn. (24). Therefore,
$R_{AD}\propto{\cal M}_A$: we expect that, on average, the strong coupling
approximation is increasingly good as ${\cal M}_A$ increases. For example,
in our $N=128^3$ simulations,
the value of $\RAD$ spans a range from about 5 to about 300, 
for $0.5\le {\cal M}_A\le 30$. In practice, however, $L_B$ is bounded below by the
resolution of the grid, leading to a flattening of the $L_B$ {\it vs} ${\cal
M}_A$ relation as ${\cal M}_A$ increases. This can be seen in Figure 3 for
${\cal M}_A\ga 10$. Thus, although
we do not accurately model MHD shocks in
which two fluid effects are significant (see Draine \& McKee 1993), 
we do resolve much of the magnetic field structures outside of shocks, 
at least in simulations with ${\cal M}_A<10$.

\section{The Simulations}

The code and basic equations it solves have been previously described in 
Nordlund, Stein \& Galsgaard (1996), Nordlund \& Galsgaard (1997),
\nocite{Nordlund+Galsgaard97mhd}
\nocite{Nordlund+96para95} 
Padoan et al. (1998), and Padoan \& Nordlund (1999). Here, we include ambipolar
drift, in the strong coupling approximation described in \S 2, for the first
time, but this has not required changes in the numerical method.

The code solves the compressible MHD equations on a 3--dimensional 
staggered mesh, 
with volume centered mass density and thermal energy, face centered 
velocity and magnetic field components, and edge centered electric 
currents and electric fields (Nordlund, Stein \& Galsgaard 1996):
\nocite{Nordlund+96para95}

\def\vv{{\bf v}}
\def\jj{{\bf j}}
\def\bb{{\bf B}}
\def\lnr{\ln\rho}
\def\div{\nabla\cdot}

\begin{equation}
\label{8}
{\partial \ln\rho \over \partial t} + \vecv \cdot \nabla\lnr = - \div \vecv,
\end{equation}

  \begin{equation}
   {\partial{\vecv} \over \partial t}
   + {\vecv\cdot\nabla\vecv}
  =
   - {P\over\rho} \nabla \ln P
   + {1\over c\rho} {\vecJ} \times {\vecB}
   + {\bf f},
  \label{9}
  \end{equation}


\begin{equation}
\label{10}
P = \rho T,
\end{equation}


\begin{eqnarray}
\label{11}
{{\partial\vecB}\over {\partial t}} = \nabla\times (\vecv\times\vecB) +
\nabla\times{{(\vecJ\times\vecB)\times\vecB}\over {c\rhoi\nuin}}.
\end{eqnarray}

\begin{equation}
\frac{4\pi}{c}\vecJ = \nabla\times\vecB,
\label{12}
\end{equation}

\noindent
plus numerical diffusion terms, and
with periodic boundary conditions. The system is forced by ${\bf f}$, an
external random driver, and all the other symbols have their usual meanings.

In calculating the dynamics we use an isothermal equation of state, as
in our previous work, even though in this paper and the next we are
concerned with the thermal equilibrium of the gas, and preliminary
results suggest a significant spread of temperatures. As long
as the motions remain highly supersonic, the dynamics should be
almost insensitive to the thermal pressure, although the equation of state
certainly affects the density contrast produced in shocks.

We use spatial derivatives accurate to 6th order, interpolation accurate 
to 5th order, and Hyman's 3rd order time stepping method (Hyman 1979).
\nocite{Hyman1979}
The code uses shock and current sheet capturing techniques to ensure
that magnetic and viscous dissipation at the smallest resolved scales 
provide the necessary dissipation paths for magnetic and kinetic energy.
As shown by Galsgaard \& Nordlund (1996, 1997), dissipation of 
\nocite{Galsgaard+Nordlund95xc,Galsgaard+Nordlund95ff}
magnetic energy in highly turbulent, compressible MHD 
plasmas occurs at a rate that
is independent of the details of the small scale dissipation. 
In ordinary hydrodynamic turbulence the corresponding property is
one of the cornerstones of Kolmogorov (1941) scaling.
\nocite{Kolmogorov41}

The initial random velocity field is generated in Fourier space,
with a normal distribution of amplitudes
and random phases, and with power only in the Fourier
components in the shell of wave--number $1\le kL/2\pi\le2$. 
In all but one of the experiments the flow is driven by an external
random force (${\bf f}$ in the momentum equation (\ref{9})).
The random force is also generated in Fourier space with a 
normal distribution, with power only in the range of  
wave--number $1\le kL/2\pi\le2$. A Helmholtz decomposition is
performed on both the initial velocity and the random force,
and only the solenoidal component is used. In order to produce a
force that varies continuously in time, we actually specify 
randomly distributed Fourier components of the time derivative
of the acceleration, and compute the acceleration as a time
integral. The time derivative of the Fourier components of the 
acceleration are regenerated at time intervals of about one dynamical time.

All our models have periodic boundary conditions. This is an efficient way
to study the evolution of MHD turbulence under internal forces, but it
precludes study of the coupling between molecular clouds and the surrounding
diffuse medium, such as, for example, outward radiation of cloud energy in
the form of MHD disturbances (Elmegreen 1985).
Since we  do not have a physically motivated model for the initial conditions,
we allow the models to relax before studying their statistical
properties, and report only results for the steady state.

For the purpose of computing the ambipolar drift heating rate, we have run
several experiments with external random driving, at an approximately
constant sonic rms Mach number, 
${\cal M}\approx 10$, with different values of the magnetic field strength, 
in a 128$^3$ numerical grid (named A1 to A7 in Table 1), and a 
decaying experiment (no external driving), with approximately 
constant value of the rms Alfv\'{e}nic Mach number, ${\cal M}_A\approx 1$, 
and magnetic field strength that decreases with time (named B1 in Table 1).
In order to perform a convergence study, we have repeated a 128$^3$ 
experiment at lower resolutions: 64$^3$, 32$^3$, and 16$^3$ (experiments
A4 in Table 1). Finally, we have repeated a 128$^3$ experiment with different 
values of the ambipolar drift parameter $a$, introduced below
(experiments A3 in Table 1). The results
of these experiments are presented in the next section, and
some of their parameters are listed in Table 1.
When we rescale the experiments to physical units (see below)
we use an average gas density $\langle n\rangle=320$~cm$^{-3}$,
according to the Larson relations (Larson 1981; Solomon et al. 1987)
and to ${\cal M}\approx 10$. Although ionization by UV photons
is generally important at this relatively low density (and likely
low extinction), we assume in the following that the gas is ionized 
by cosmic rays. We do so because, while the volume averaged gas
density is relatively low, most of the gas is actually at densities
larger than ten times the mean density, due to the very intermittent
density distribution (Padoan et al. 1998; Padoan \& Nordlund 1999). 
In numerical experiments with ${\cal M}\approx 10$
and numerical mesh size of 128$^3$, the complex system of interacting shocks
generates a very large density contrast of almost six orders of 
magnitude (0.1~cm$^{-3}\la n\la 10^5$~cm$^{-3}$). However, the ambipolar
drift heating rate computed in this work and expressed in eqn. (\ref{eqn_fit})
can be rescaled to any value of the average gas density, and it is
increasingly accurate for increasing average density, because of the 
assumption of cosmic ray ionization.

Cosmic rays ionization results in the following
expression for the fractional ionization $x_i$:

\begin{equation}
\label{xi}
x_i \equiv \frac{K_i}{n_n^{1/2}} 
\end{equation}
where $n_n$ is the number density of neutrals, and 
$K_i$ is of order $10^{-5}$~cm$^{-3/2}$ (see the discussion 
of $K_i$ in McKee et al. 1993). 
We have not included
photoionization by the ambient radiation field, although it is
important at low extinction (McKee 1989). Equation (13) underestimates the
ionization, and hence overestimates the degree of ambipolar drift, in the
low density gas, but our main interest is in the high density regions,
which contain most of the mass.

In order to explain our parameterization of ambipolar drift, we must introduce
our natural units and nondimensionalized equations. We denote quantities
defining a unit by a subscript $u$ and
dimensionless quantities by tildes,
that is $\tilde{\vecB}=\vecB /B_u$, 
$\tilde{\vecv}=\vecv /v_u$, $\tilde{\vecJ}=\tilde{\nabla}\times\tilde{\vecB}$,
$\tilde{\nabla}=L_u\nabla$, and $\tilde{n}=n /n_u$ . The code is written 
so that
the units are:

\begin{equation}
\label{bu}
B_u=c_S(4\pi\langle \rho\rangle)^{1/2}
\end{equation}

\begin{equation}
\label{vu}
v_u=c_S
\end{equation}

\begin{equation}
\label{nu}
n_u=\langle n\rangle
\end{equation}

\begin{equation}
\label{nablau}
L_u=L/N
\end{equation}
where $c_S$ is the sound speed, $L$ the physical size of the 
simulation box, and $N$ the linear resolution of the numerical
box.

The ambipolar drift velocity is written as

\begin{equation}
\label{vDcode}
\tilde{\vecv}_D=a \tilde{n}^{-3/2}{\tilde{\vecJ}\times\tilde{\vecB}}
\end{equation}

 The ambipolar drift parameter $a$ can be computed from
the expression $\tilde{\vecv}_D=\vecvD /v_u$, and using equation
(\ref{vDs}) and (\ref{vDcode}), together with (\ref{bu}), 
(\ref{vu}),(\ref{nu}) and (\ref{nablau}). The result is:

\begin{equation}
\label{a}
a=0.28\left(\frac{T}{10K}\right)^{1/2}\left(\frac{N}{128}\right)
       \left(\frac{L}{10\,{\mathrm pc}}\right)^{-1}
       \left(\frac{\langle n\rangle}{200\,{\mathrm cm^{-3}}}\right)^{-1/2}
       \left(\frac{K_i}{10^{-5}\,{\mathrm cm^{-3/2}}}\right)^{-1}
       \left(\frac{<\sigma v>}{2\times 10^{-9}\,{\mathrm cm^3\,s^{-1}}} \right)^{-1}.
\end{equation}
The value of the parameter $a$ that we use in most of the present
experiments is $a=0.3$ in 128$^3$ runs, and appropriately
rescaled for different resolutions. Using equation (\ref{a}),
$a=0.3$ corresponds to physical conditions in typical molecular
clouds with temperature $T=10$~K, size $L=10$~pc, and
volume averaged number density $\langle n\rangle=200$~cm$^{-3}$.
Using Larson's relation between size and density
(Larson 1981; Solomon et al 1987), $\langle n\rangle\propto L^{-1}$,
the dependence of $a$ on the cloud size would be
$a\propto L^{-1/2}$. As an example, a cloud or cloud core
with $L=1$~pc and $\langle n\rangle=2000$~cm$^{-3}$,
would require a value of $a=0.88$, while a giant molecular cloud
complex with $L=50$~pc should be modeled with $a=0.12$. 
Physical conditions in molecular clouds with sizes in the 
range 1--50~pc can therefore be reproduced numerically with 
values of $a$ within a factor of less than 3 from the value
$a=0.3$ (at $N=128^3$ resolution) adopted in most of
the numerical experiments used in this work. 
However, the dependence of the ambipolar drift heating rate on
the parameter $a$ has been computed (see Figure 5),
by varying the value of $a$ in the simulations
in the range $0.15\le a \le 1.0$, that corresponds to
cloud size $0.5$~pc$\la L\la 50$~pc.

\section{The Ambipolar Drift Heating Rate}

Our results are based on
a variety of both driven and decaying cloud models, over a range
of parameters and at different resolutions,
as described in the previous section. The heating rates that we report in
the driven models all apply to the steady state, which is reached after
approximately 3-4 dynamical times. However, because all quantities
fluctuate slightly with time with respect to their steady state values, we
typically enter values from several adjacent time slices of each simulation
on the figures which follow.

Although, as we show later in this section, the heating rate varies
considerably from point to point, the volume averaged heating rate $\langle
H_{AD}\rangle$ scales with the properties of the system in a remarkably simple
way. We define the mean of a quantity $Q$, $\langle Q\rangle$, by
\begin{equation}
\langle Q\rangle\equiv N^{-1}\sum_{ijk}Q_{ijk},
\end{equation}
where $ijk$ indexes a grid point. From eqn. (6),
\begin{equation}
\langle H_{AD}\rangle = N^{-1}\sum_{ijk}(\rho_i\nu_{in}\mid\vecvD\mid^2)_{ijk},
\end{equation}
where $\vecvD$ is computed from eqn. (4).

We have performed a convergence study of the dependence of $\langle H_{AD}
\rangle$ on numerical resolution. Figure 1 shows $\langle H_{AD}\rangle$ for
four simulations with $N= 16^3$, $32^3$, $64^3$, and $128^3$, and shows clear
evidence for convergence, probably to better than a factor of 2, of the value
found for the highest resolution. 

Studies at high spatial resolution have shown that ambipolar drift can
steepen the current density profile to the point that it virtually becomes
singular (Brandenburg \& Zweibel 1994, 1995, Mac Low et al 1995, Zweibel \&
Brandenburg 1997, Mac Low \& Smith 1997). However, the contribution of such
structures to the total heating rate is small, as shown by the following
argument. The magnetic field varies with
position $x$ across a sheet as $B\sim x^{1/3}$, so $J$ varies as $x^{-2/3}$.
From eqns. (4) and (6), $H_{AD}\propto x^{-2/3}$, which is singular, but
integrable: the integrated value of $H_{AD}$ 
depends on the width $x_s$ of the sheet as $x_s^{1/3}$. 
Therefore,
these very thin structures contribute little to the heating rate - even less
if one takes into account two fluid effects, which resolve the singularity.

Assuming that the heating rate is nearly converged, the
scaling law we find for $\langle H_{AD}\rangle$ depends on the mean
magnetic field magnitude $\langle\mid B\mid\rangle$
\begin{equation}
\langle\mid B\mid\rangle\equiv\langle(B^2)^{1/2}\rangle,
\end{equation}
the rms velocity $v_{rms}$
\begin{equation}
v_{rms}\equiv\langle v^2\rangle^{1/2},
\end{equation}
and the Alfven Mach number in the evolved state
\begin{equation}
{\cal M}_A\equiv{{v_{rms}(4\pi\langle\rho\rangle)^{1/2}}\over {\langle\mid B\mid
\rangle}}.
\end{equation}
Figure 2 shows $\langle H_{AD}\rangle/{\cal M}_A^2$ plotted versus $\langle
\mid B\mid\rangle$ for a set of simulations with constant sonic Mach number
${\cal M}=10$. Thus, ${\cal M}_A\propto\langle\mid B\mid\rangle^{-1}$ with a
unique constant of proportionality, so we plot ${\cal M}_A$ on the top axis
of the figure. In all of these runs, the ambipolar drift parameter $a$
defined in eqn. (19) is fixed at 0.3 for $N=128^3$. According to 
the Larson relations, this 
model corresponds to a cloud
with $L\sim 6$ pc, $\langle n\rangle\sim 320$ cm$^{-3}$.

Figure 2 shows a remarkably tight correlation between $\langle\mid B\mid\rangle
$ and $\langle H_{AD}\rangle$. The simulations are fit by the relation
\begin{equation}
\label{eqn_fit}
\langle H_{AD}\rangle = 3.0\times 10^{-24}\Bigg({{\langle\mid B\mid\rangle
}\over {10\mu G}}\Bigg)^4\Bigg({{{\cal M}_A}\over {5}}\Bigg)^2\Bigg({{\langle
n\rangle}\over {320 cm^{-3}}}\Bigg)^{-3/2} {\mathrm erg\,cm^{-3}\,s^{-1}},
\end{equation}
which is shown as a solid line on Figure 2.

The cosmic ray heating rate $H_{cr}$ corresponding to a primary cosmic ray
ionization rate $\zeta = 10^{-17}$~s$^{-1}$ is also plotted in Figure 2. 
Ambipolar drift heating exceeds cosmic ray heating unless the magnetic field
is less than about 3~$\mu$~G. However, the value 
$\langle n\rangle=320$~cm$^{-3}$ has been used in 
Figure 2; for increasing values of the average gas density, 
the cosmic ray heating rate increases, since it is proportional
to the gas density. The cosmic ray heating rate is also found to
vary considerably from cloud to cloud, and could span the range
of values $10^{-16}$~s$^{-1}\la \zeta \la 10^{-18}$~s$^{-1}$
(Caselli et al. 1998; Williams et al. 1998).

Equation (6) is only compatible with eqn. (\ref{eqn_fit}) if
\begin{equation}
\label{eqn_LB}
\langle L_B\rangle \equiv\langle\Bigg({{c^2 B^2}\over {(4\pi J)^2}}\Bigg)^{1/2}\rangle\propto{\cal M}_A^{-1}.
\end{equation}
Figure 3 shows that this is in fact the case, unless ${\cal M}_A>10$. For
${\cal M}_A>10$, although the value of the volume averaged magnetic
length scale, $\langle L_B\rangle$, is still a bit larger than the numerical grid scale,
local values of $L_B$ are so small that in many locations
$L_B$ is not resolved by
the 128$^3$ numerical mesh. It is possible that $\langle L_B\rangle\propto {\cal M}_A^{-1}$
also for ${\cal M}_A>10$, but this cannot be shown in numerical runs 
with the present resolution. 

Qualitatively, it is not surprising that $\langle L_B\rangle$ decreases with increasing 
${\cal M}_A$,
or increasingly dominant kinetic energy. The larger 
${\cal M}_A$ is, the more the flow
can bend the field, leading to more tangling and smaller $L_B$. 
What perhaps is surprising is that in cases in which the
field is initially weak, it is not amplified up to equipartition: if it were,
there would be no points in Figure 3 with ${\cal M}_A > 1$. A number of factors
can lead to saturation of the field below equipartition. The weaker the field
is initially, the more it must be stretched and tangled to amplify it up to
equipartition. The more tangled the field becomes, the more subject it is to
numerical and ambipolar diffusion. It has also been shown (Brandenburg \&
Zweibel 1994, Brandenburg et al 1995, Zweibel \& Brandenburg 1997) that 
ambipolar drift drives the magnetic field to a nearly force free state. Since
the flow cannot do work on a force free field, amplification ceases once the
force free state is reached. And finally, the magnetic field becomes quite
spatially intermittent (see Figure 7), and locally strong Lorentz forces might
feed back on the flow and quench the growth of the field even before ambipolar
drift has had  enough time to act.

The foregoing results on $\langle H_{AD}\rangle$ and the behavior of $\langle L_B\rangle$ are
shown in another form in Figure 4, which is based on a decaying run with no
driving and no mean magnetic field. The magnetic and kinetic energies are
initially in equipartition, and remain so as they decrease with time. The
magnetic field strength shown on the abscissa parameterizes time from late to
early. Figure 4 shows that the sonic Mach number ${\cal M}$ decays while $\langle L_B\rangle$
and ${\cal M}_A$ remain roughly constant. Despite fluctuations, $\langle H_{AD}
\rangle$ is proportional to $\langle\mid B\mid\rangle^4$ as the 
magnetic field decays
by an order of magnitude.

We have also studied the dependence of $\langle H_{AD}\rangle$ on the ambipolar
drift parameter $a$. Taken at face value, eqn. (6) predicts $\langle H_{AD}
\rangle\propto a$. But this reasoning does not account for the dependence
of the magnetic field properties on $a$. As $a$ increases, the increase in
magnetic diffusion decreases the efficiency with which the field is amplified,
and increases $L_B$. 
The net result is that although $\langle H_{AD}\rangle$ is positively
correlated with $a$, the dependence is weaker than linear over the range
of $a$ which we have examined. $H_{AD}$ and $L_B$ are plotted versus $a$
in Figure 5. The plot of $\langle H_{AD}\rangle$ versus $a$ is reasonably well fit by 
$\langle H_{AD}\rangle\propto a^{0.6}$ over the limited but physically reasonable 
range $0.15\le a\le 1.0$. The increase in $\langle L_B\rangle$ by roughly a factor of 
2 over this range of $a$ accounts for the increase of $\langle H_AD\rangle$ by roughly 
a factor of 4 instead of a factor of 7, as would occur for linear dependence.

Ambipolar drift makes a significant contribution to the rate at which the
turbulence decays. Writing the kinetic energy $E_K$ as
\begin{equation}
E_K = {{1}\over {2}}\langle\rho\rangle v_{rms}^2={\cal M}_A^2{{\langle\mid B\mid
\rangle^2}\over {8\pi}}
\end{equation}
and using eqn. (\ref{eqn_fit}) we find the decay time $\tau_{AD}$
\begin{equation}
\tau_{AD}\equiv{{E_K}\over {\langle H_{AD}\rangle}}=1.3\times 10^6\Bigg({{
\langle\mid B\mid\rangle}\over {10\,{\mathrm \mu G}}}\Bigg)^{-2}\Bigg({{\langle n\rangle}
\over {320\,{\mathrm cm^{-3}}}}\Bigg)^{3/2}\,{\mathrm yr}.
\end{equation}
For comparison, in these models the dynamical time scale $L/v_{rms}$ 
is about $2.4
\times 10^6$ yr. Thus, as has been argued elsewhere, ambipolar drift is an
efficient mechanism for the dissipation of turbulent energy in molecular
clouds (Zweibel \& Josafatsson 1983, Elmegreen 1985).

Recently, it has been shown that supersonic hydrodynamic and hydromagnetic
turbulence both decay as $E_K\propto t^{-\eta}$, with $\eta\sim 1$ (Mac Low
et al 1998, Mac Low 1999). This decay law implies
\begin{equation}
{{dE_K}\over {dt}}\propto -E_K^{2}.
\end{equation}
In these models, dissipation occurs primarily in shocks and through the
generation and numerical dissipation of short wavelength Alfven waves; there
is no ambipolar drift.

Ambipolar drift actually leads to a similar decay law: $\langle H_{AD}\rangle
\propto\langle\mid B\mid\rangle^2 v_{rms}^2$, 
which can be written as $E_K^2$ since 
the ratio of magnetic to kinetic energy is fixed in these decaying models,
as shown in Figure 4.
Does it matter whether energy is lost by ion-neutral friction or in shocks?
From an observational point of view it does matter, since the peak temperatures
and spatial distribution of the heating in the two cases can be quite
different. We will explore this issue further in a study of thermal equilibrium
in molecular clouds heated by ambipolar drift (Juvela et al. 2000). 

Having established the mean properties, we now describe the
pointwise properties of ambipolar drift heating.
Figure 6 is a scatter plot, in dimensional units, of $H_{AD}$ versus $4\pi B^2
v^2\rho^{-1/2}$ for points at which the density satisfies the condition
$n > 2\langle n\rangle$. We choose to look only at relatively dense regions
since we probably overestimate the ambipolar drift 
rate at low densities due to our choice of ionization law, eqn. (13). 
Moreover, we have plotted only a
randomly selected subset of the points that satisfy the density condition, 
because there are so many points in the simulation that plotting 
all of them would 
produce a completely saturated, solid black clot. 
The solid line in Figure 6, which is
the mean heating 
relation given in eqn. (\ref{eqn_fit}), is an approximate fit to the
data, but there is a dispersion of a few orders of magnitude in the 
heating rates from point to point.

Figure 7 shows images of two--dimensional slices of $H_{AD}$ (left panel) and $\rho$ (right
panel), both plotted on a grey scale corresponding to (magnitude)$^{0.1}$.
Figure 7 shows that both the heating and density distributions are highly
inhomogeneous and filamentary, and that high density features are generally
regions of strong heating. Bearing in mind that $H_{AD}\propto\rho^{-3/2}$,
we see that the strong heating in high density regions must be a consequence
of the $B-\rho$ correlation and also the small values of $L_B$ associated with
thin, magnetized filaments. 

A close examination of the heating and density distributions in Figure 7 shows
thin local minima in the heating sandwiched between adjacent ribbons of strong
heating. This phenomenon is seen more easily in Figure 8, which shows ambipolar
drift heating and cosmic ray heating on a cut taken through the images shown
in Figure 7. We have chosen to plot $H_{CR}$, which is linear in $\rho$, as a
proxy for the density because this makes it easy to compare the relative
magnitudes of ambipolar drift and cosmic ray heating. It is clear that density
peaks are often flanked by peaks of $H_{AD}$. The reason for this is not hard
to 
understand. The magnetic field tends to be aligned with the density filaments,
due to the strong compression which formed the filamanets. 
Within the filaments, the drift velocity ${\bf v}_D
\propto -\nabla B^2\propto\nabla\rho$. According to eqn. (6), the heating
reaches a minimum at the center of the filament and peaks off center where
the density gradient is largest. Although the correlation between $L_B$ and
$n$ shows considerable scatter, there is substantial correspondence between
the most pronounced features in each, as shown in Figure 9.

The mean values of $H_{CR}$ and $H_{AD}$ are also plotted in Figure 8. In this
model, with $<|B|>\approx 6 \mu$~G, ambipolar drift heating 
is on average almost 4 times 
larger than cosmic ray heating.

\section{Discussion and Conclusions}

It was suggested some time ago that ambipolar drift, whether arising from the
diffusion of a large scale magnetic field or from damping small scale
turbulence, could be an important heating mechanism for molecular clouds
(Scalo 1977, Goldsmith \& Langer 1978, Zweibel \& Josafatsson 1983). 
It was also
pointed out in these papers that the observed low temperatures of
molecular clouds implies bounds on the magnitude of the magnetic field, and/or
the amplitude of the turbulence.

In practice, however, the magnitude of the ambipolar drift heating $H_{AD}$
has been difficult to assess. Simple arguments show that $H_{AD}$ should scale
as $B^4/(L_B^2\rho\rho_i)$ (or $B^4/(L_B^2\rho^{3/2})$ for a cloud ionized by
cosmic rays). The extreme sensitivity of $H_{AD}$ to $B$, which at best can be
measured only along the line of sight, and to the magnetic length scale $L_B$,
which cannot be measured at all, make it difficult to estimate $H_{AD}$
reliably to better than an order of magnitude. This has made ambipolar drift
heating appear less attractive than mechanisms which can be reliably evaluated,
such as heating by low energy cosmic rays. It has also 
seriously weakened the constraints
on magnetic field properties which arise from thermal balance calculations.
The further difficulty of calculating radiative cooling rates which account
self consistently for the turbulent structure of the cloud, including the
effects of turbulence on cloud chemistry, has only compounded the problem.

For several reasons, the role of ambipolar drift heating in the thermal
equilibrium of clouds is becoming more open to assessment, and hence more
interesting. Reliable maps of temperature, density, magnetic field, and
velocity structure are now available for more clouds than ever before,
permitting correlation studies of cloud temperature with other properties.
Detailed numerical models of turbulent molecular clouds, and improved
calculations of cooling rates and cloud chemistry, are now feasible.
Thermal balance calculations which include ambipolar drift heating, compared
with observations, can now lead to meaningful diagnostics of magnetic fields
and dynamical properties of molecular clouds.

In this paper we have used simulations of turbulent, magnetized molecular
clouds to study the properties of heating by ambipolar drift. The models are
self consistent in the sense that we include ambipolar drift in calculating
the dynamics of the cloud and the evolution of the magnetic field. We have
found that a realistic amount of ambipolar drift in a simulation with the
highest numerical resolution practical for us ($N=128^3$ mesh points) imparts
a physical as opposed to numerical diffusivity to the cloud which makes it
just possible to capture all of the relevant length scales. The models can
be scaled to approximate the sizes, densities, velocity dispersions, and
magnetic field strengths of observed clouds. As our simulations do not include
self gravity, we focus on ambipolar drift heating due to turbulence rather than
due to the systematic redistribution of a large scale field in quiescent,
gravitationally contracting structures.

The main result of our paper is an empirical formula for the volume averaged
heating rate, $\langle H_{AD}\rangle$, given in eqn. (\ref{eqn_fit}) : $\langle
H_{AD}\rangle\propto\langle\mid B\mid\rangle^4{\cal M}_A^2/\rho\rho_i$, or
$\langle H_{AD}\rangle\propto\langle\mid B\mid\rangle^2 v_{rms}^2/\rho_i$. As
shown in Figure 2, the fit is very good, and comes about because the
magnetic length scale $L_B$ is inversely proportional to ${\cal M}_A$ (Figure
3). Our scaling law for $\langle H_{AD}\rangle$ makes it possible to estimate
the ambipolar drift heating rate in individual clouds much more accurately
than before: the uncertainty in the value of $B$ is only squared, not raised
to the fourth power, and $L_B$ is replaced by the more readily measurable
rms velocity, or line width. 

Furthermore, $\langle H_{AD}\rangle$ turns out to be interestingly large,
exceeding, for moderately large field strength, the mean cosmic ray heating
rates $H_{CR}$. And, the turbulent dissipation times associated with ambipolar
drift heating are of order the dynamical crossing times in our models,
suggesting, as argued elsewhere, that the turbulence in molecular clouds must
be driven continuously.

In some respects, however, the conventional wisdom that $H_{AD}$ is difficult to
estimate is borne out by its large range in value 
from point to point within any particular model.
Despite the excellent correlation of the global heating rate with averaged
quantities, a scatter plot of $H_{AD}$ versus the pointwise values of the
same quantities which go into the mean relation eqn. (25)
shows dispersion of a few orders of magnitude about the mean, as shown in
Figure 6,
(although the relation based on global averages is an approximate fit to the centroid
of the distribution). In models with a fairly weak magnetic field and a tight
$B-\rho$ relation, the correspondence between density and $H_{AD}$ is quite
good, even down to individual features (Figure 7).

We are presently studying the thermal equilibrium problem with a Monte Carlo
treatment of radiative transfer. For the present, however, we note that 
standard calculations of radiative cooling rates in the temperature and density
regimes of interest to us here ($T\sim 10 K- 40 K$, $n\sim 10^3 - 10^4$)
scale with temperature as $T^{2.2-2.7}$ (Goldsmith \& Langer 1978, Neufeld,
Lepp, \& Melnick 1995). 
Therefore, if ambipolar drift is the
dominant heating mechanism, there should be a relation 
with temperature and line width of the form $T\propto \Delta v^{.74-.91}$.
The exponent of this relation could be reduced, due to the dependence
of the cooling rate on the velocity gradient in the gas. 
In fact, a positive
correlation, but with a somewhat shallower slope, has recently been reported
(Jijina, Myers, \& Adams 1999).
In general, however, we would expect that any model in which the dissipation
of turbulence contributes to heating would produce a positive $T-\Delta v$
correlation.

\acknowledgements 

We thank Ted Bergin, Alyssa Goodman, Mika Juvela and Phil Myers 
for reading the manuscript and providing useful comments.
We are happy to acknowledge support by NSF Grant
AST 9800616 and NASA Grant NAG5-4063 to the University of Colorado,
and support by the Danish National Research Foundation
through its establishment of the Theoretical Astrophysics Center.

\clearpage
\centerline{\bf FIGURE AND TABLE CAPTIONS}

{\bf Table 1:} Volume and time averaged parameters of the numerical experiments.
In the case of the decaying experiment B1, the initial values of $\langle |B|\rangle$
and  ${\cal M}$ are given, instead of their time averaged values.

\begin{figure}
\centerline{\epsfxsize=15cm \epsfbox{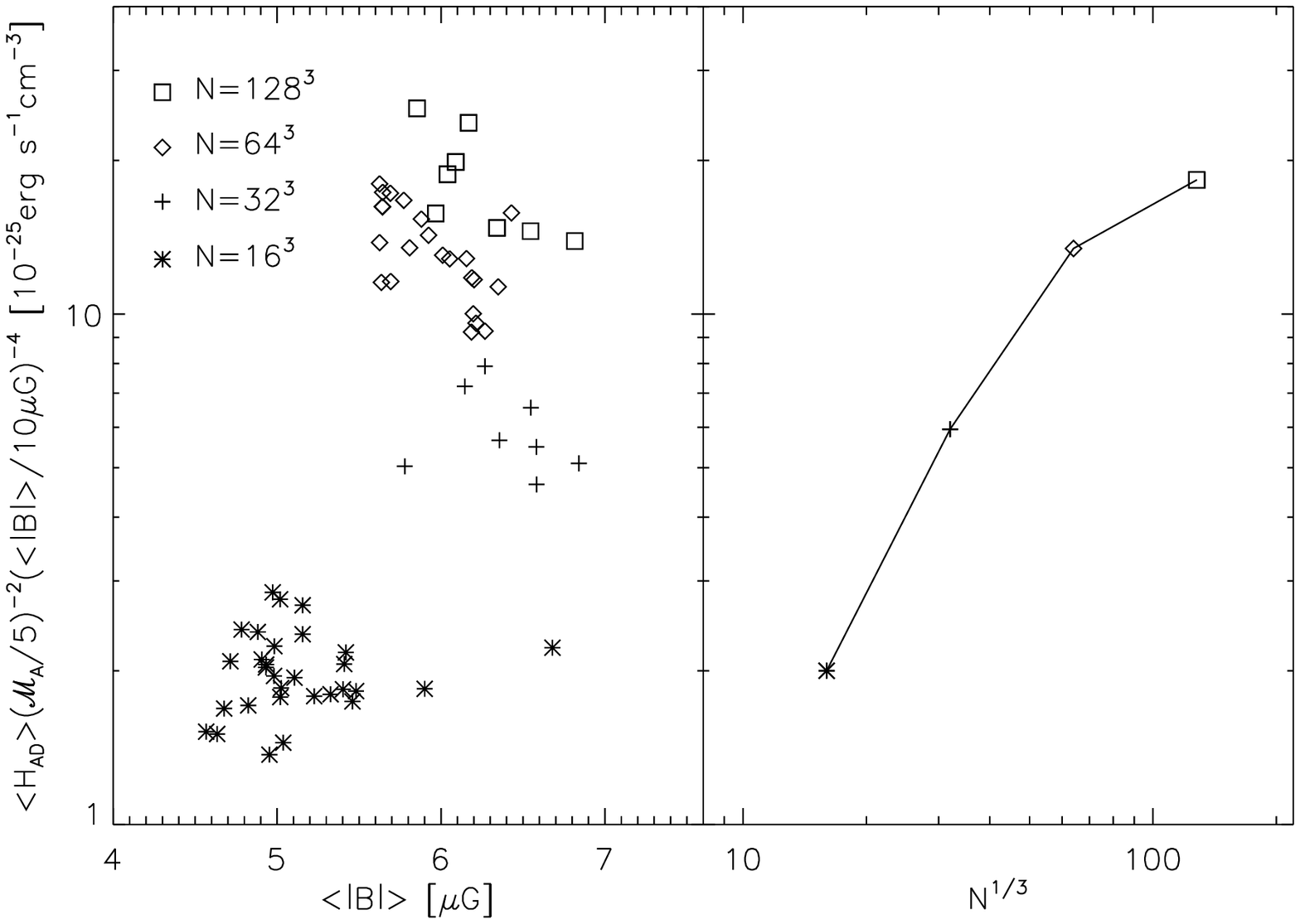}}
\caption[]{}
\label{fig1}
\end{figure}

{\bf Figure \ref{fig1}:} Left panel: Volume averaged ambipolar heating rate
per unit volume as a function of the average magnetic field strength,
in snapshots from simulations with different size of the numerical
mesh. The heating rate has been divided by its expected value
(see \S 4) in a model with ${\cal M}_A=5$ and 
$\langle |B|\rangle=10$~$\mu$G. Right panel: Heating rate averaged
over the different snapshots used in the left panel, versus the
linear size of the computational mesh. The heating rate shows 
a clear trend to convergence.\\

\begin{figure}
\centerline{\epsfxsize=15cm \epsfbox{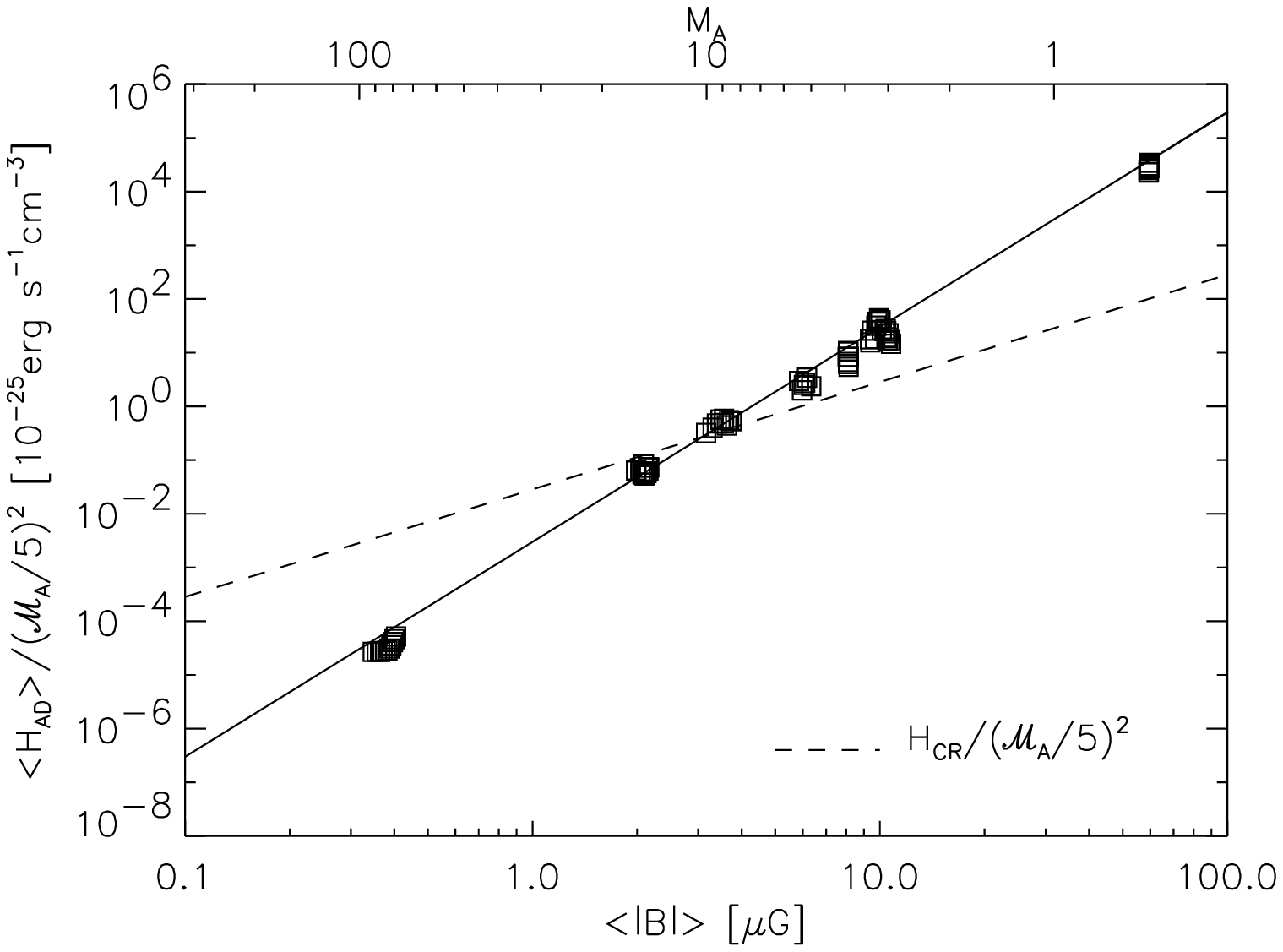}}
\caption[]{}
\label{fig2}
\end{figure}

{\bf Figure \ref{fig2}:} Volume averaged ambipolar drift heating rate,
divided by the square of the rms Alfv\'{e}nic Mach number, versus
the averaged magnetic field strength, in randomly driven 
128$^3$ simulations, with
roughly constant ordinary rms Mach number, $M\approx10$, 
and $\langle n\rangle =320$~cm$^{-3}$. The continuous
line shows a $\langle |B|\rangle^4$ dependence, while the dashed line
is the cosmic ray heating rate per unit volume, also divided by
${\cal M}_A^2$.  \\

\begin{figure}
\centerline{\epsfxsize=15cm \epsfbox{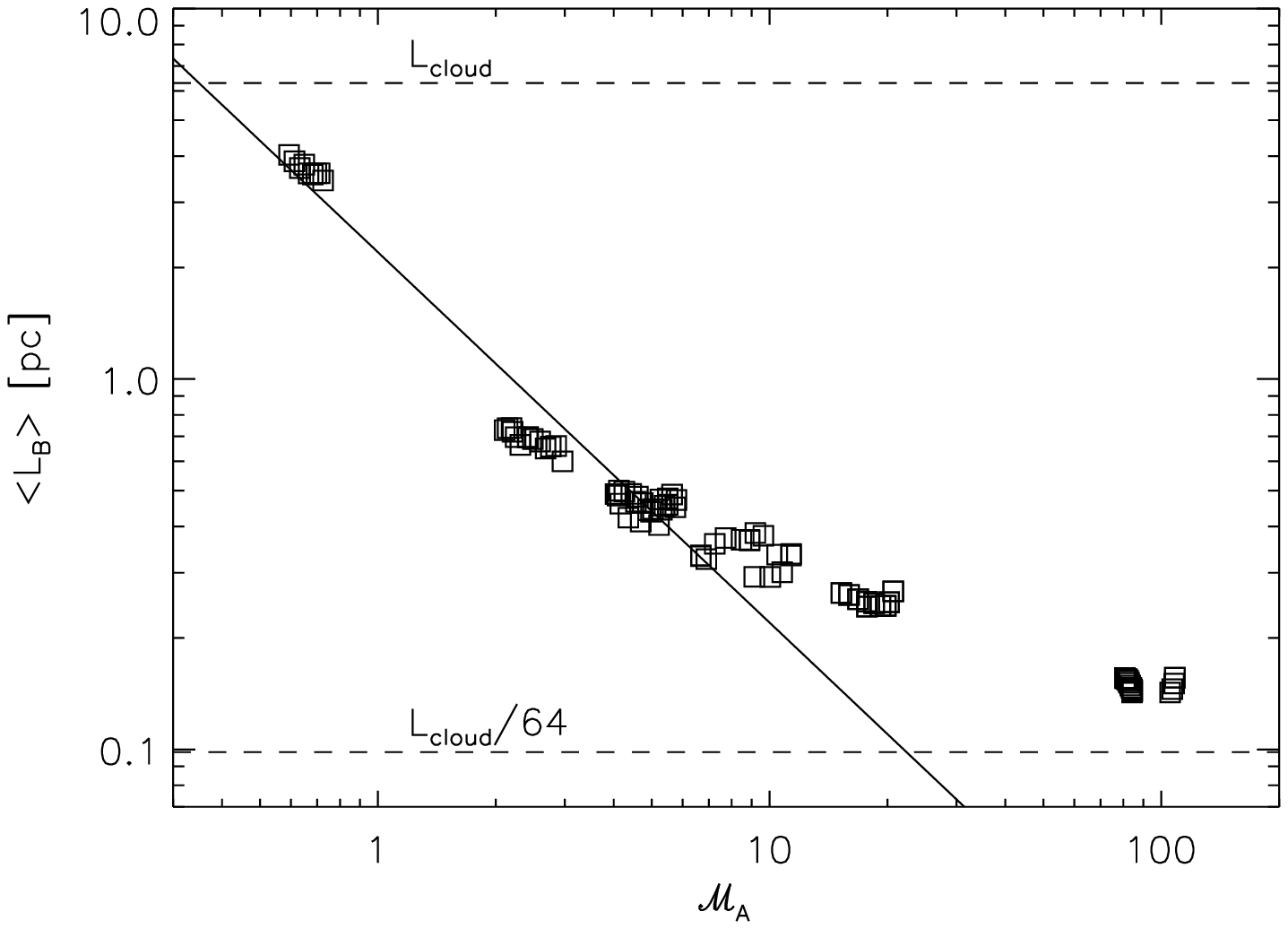}}
\caption[]{}
\label{fig3}
\end{figure}

{\bf Figure \ref{fig3}:}  Magnetic length scale (see text) versus
the rms Alfv\'{e}nic Mach number of the flow, in 128$^3$ simulations.
The upper dashed line marks the physical size of the
simulation box, and the lower dashed line the physical size that 
corresponds to the numerical resolution (taken as two grid cells).
The continuous line shows a ${\cal M}_A^{-1}$ dependence. $L_B$ is
roughly proportional to ${\cal M}_A^{-1}$, in a range of values of ${\cal M}_A$
typical of conditions in molecular clouds, while it departs from
that dependence around ${\cal M}_A\approx 10$, where $L_B$ approaches
the limit of the numerical resolution. \\

\begin{figure}
\centerline{\epsfxsize=15cm \epsfbox{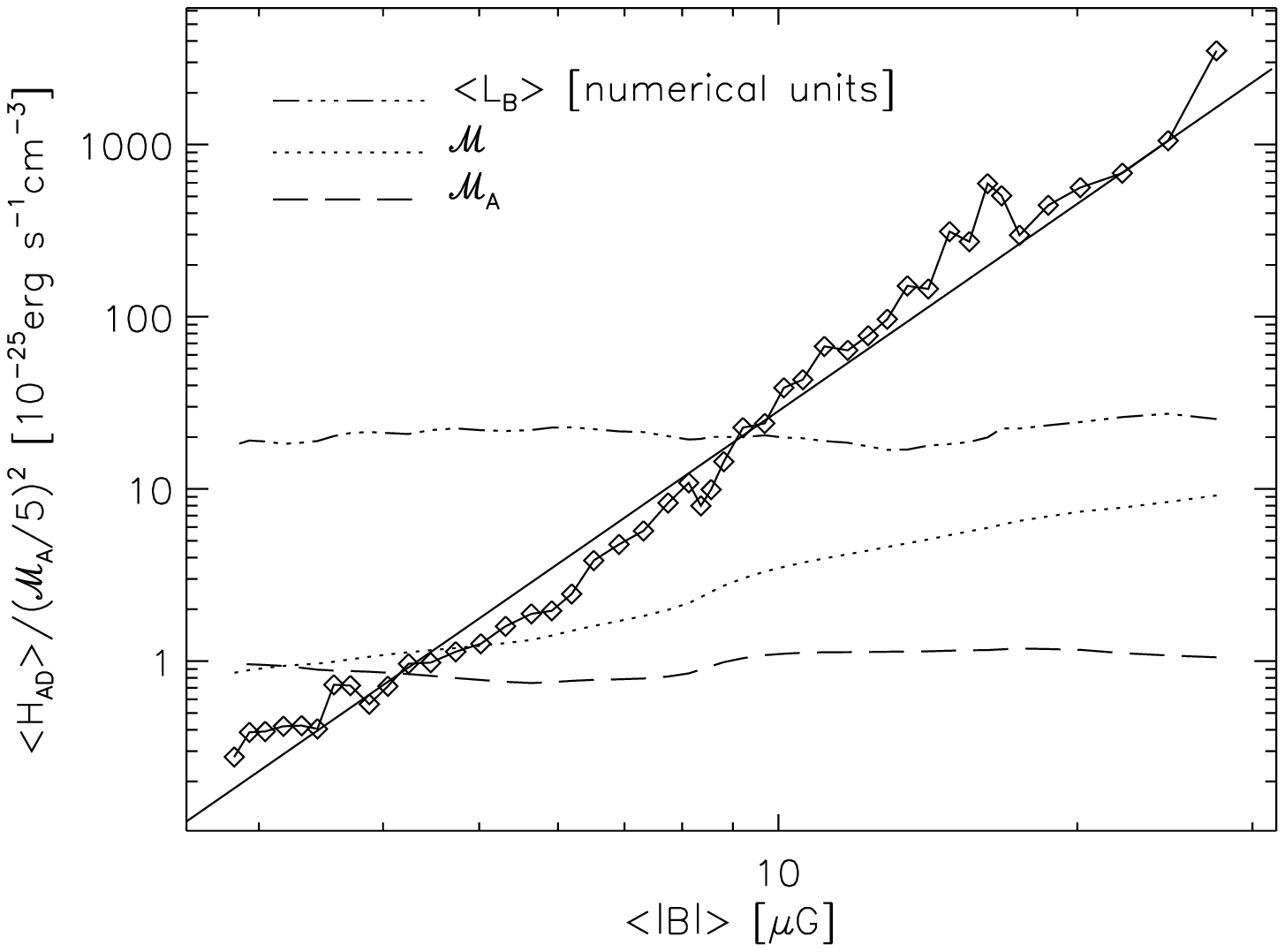}}
\caption[]{}
\label{fig4}
\end{figure}

{\bf Figure \ref{fig4}:} Volume averaged ambipolar drift heating rate
per unit volume,
divided by the square of the rms Alfv\'{e}nic Mach number, versus
the averaged magnetic field strength, in a decaying 128$^3$ simulation.
The continuous line shows a $\langle |B|\rangle^4$ dependence.
Kinetic and magnetic energy decay at approximately the same rate, and
so the flow is in approximate equipartition at all times, ${\cal M}_A\approx1$.
Also the magnetic length scale $L_B$ is approximately constant, which proves
that it depends only on ${\cal M}_A$.  \\

\begin{figure}
\centerline{\epsfxsize=15cm \epsfbox{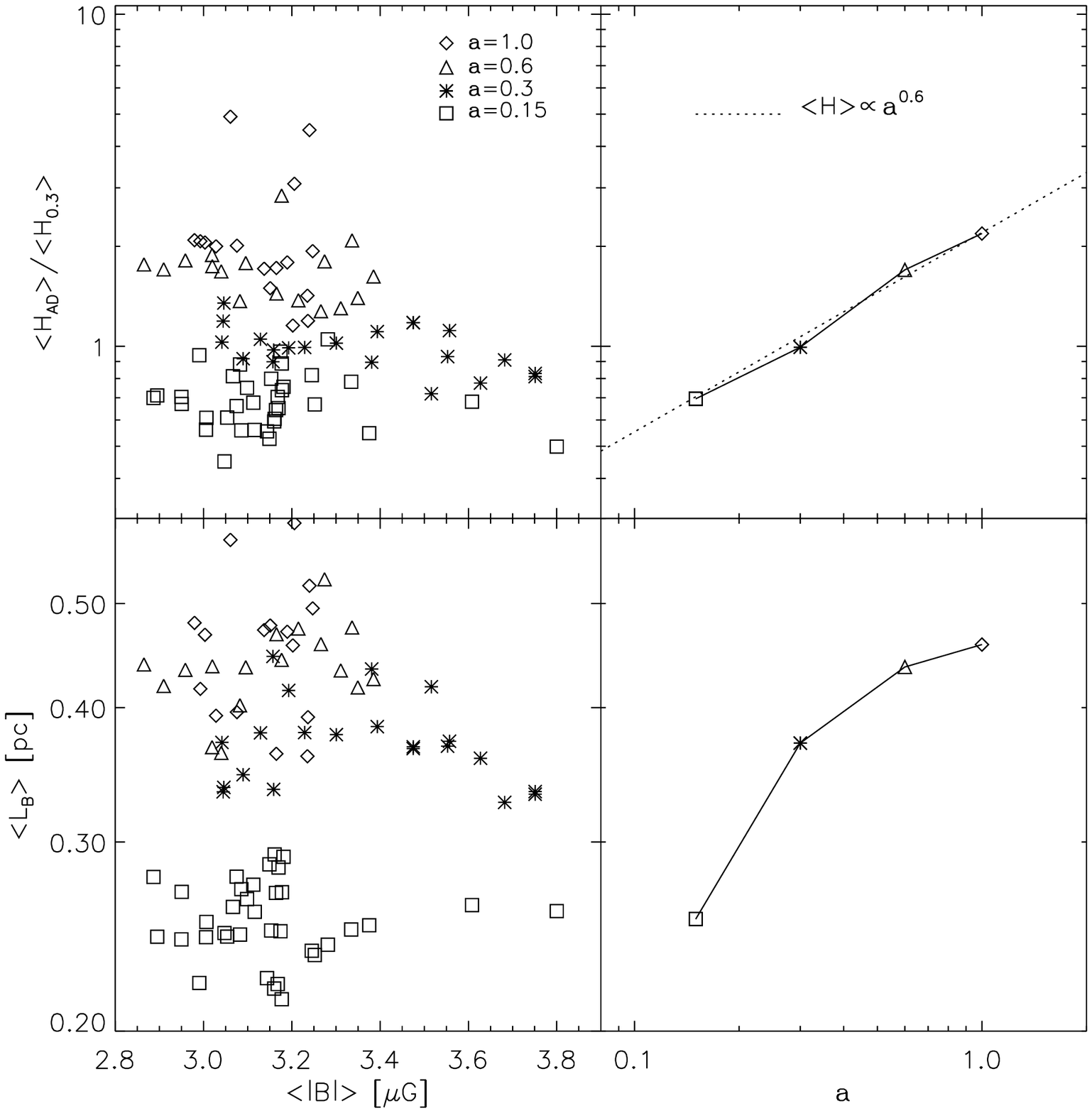}}
\caption[]{}
\label{fig5}
\end{figure}

{\bf Figure \ref{fig5}:} Upper panels: Volume averaged ambipolar drift 
heating rate per unit volume, divided by its expected value for $a=0.3$, 
according to eqn. (\ref{eqn_fit}). Different symbols represent 
different values of the ambipolar diffusion parameter, $a$. Lower 
panels: Magnetic lengthscale for different values of the ambipolar
diffusion parameter. \\

\begin{figure}
\centerline{\epsfxsize=15cm \epsfbox{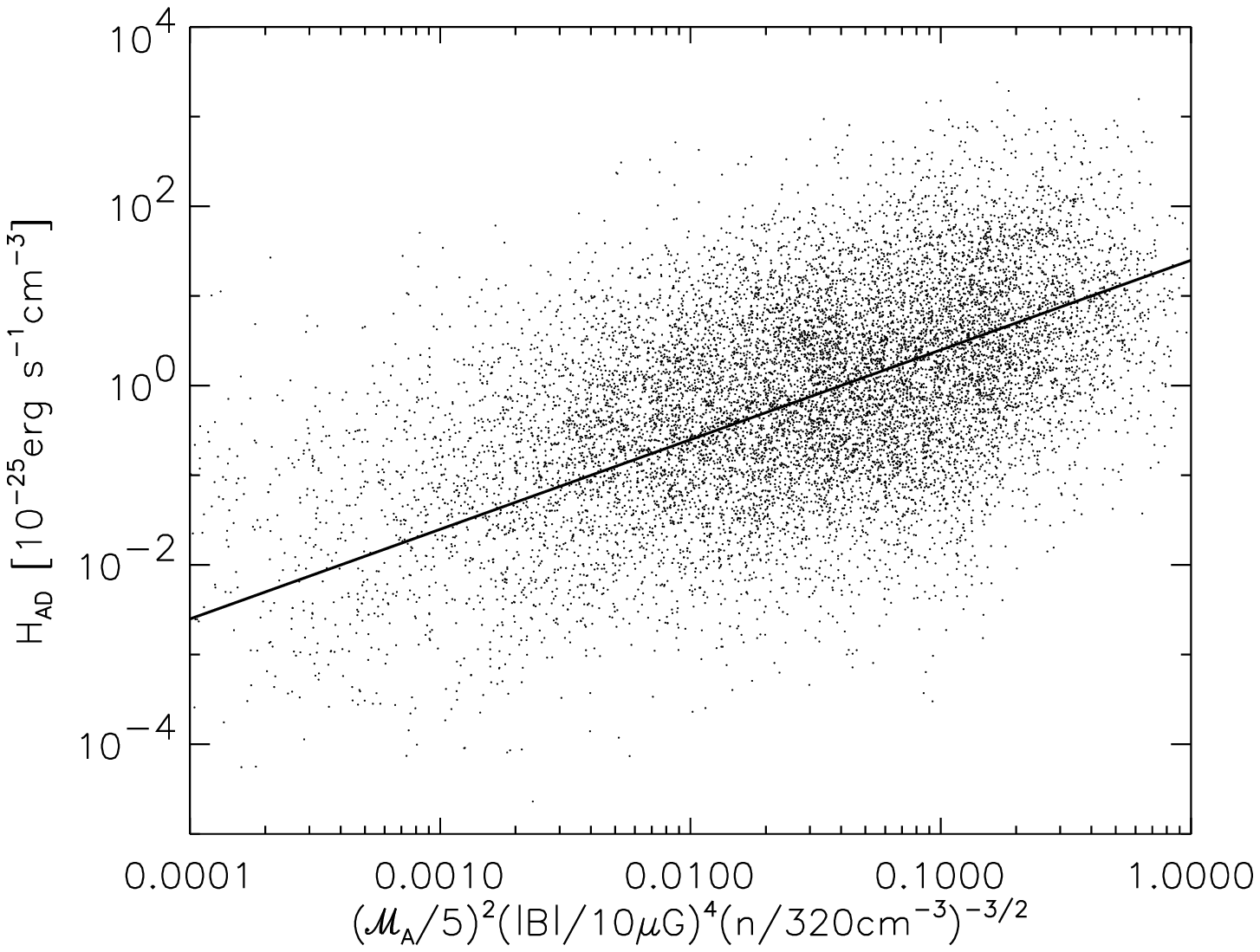}}
\caption[]{}
\label{fig6}
\end{figure}

{\bf Figure \ref{fig6}:} Local value of the ambipolar drift heating
rate per unit volume, versus its expected average value for the average
parameters used in the simulation: ${\cal M}_A=5$,
$\langle |B|\rangle=10$~$\mu$G, and $\langle n\rangle=320$~cm$^{-3}$.
The local heating rate spans a range of values of about 8 orders of
magnitude, but it also follows the value expected for 
the volume averaged heating rate, shown by the continuous line. 
Only points with $n>2\langle n\rangle$ have been used in the scatter
plot (see text). \\

\begin{figure}
\centerline{\epsfxsize=15cm \epsfbox{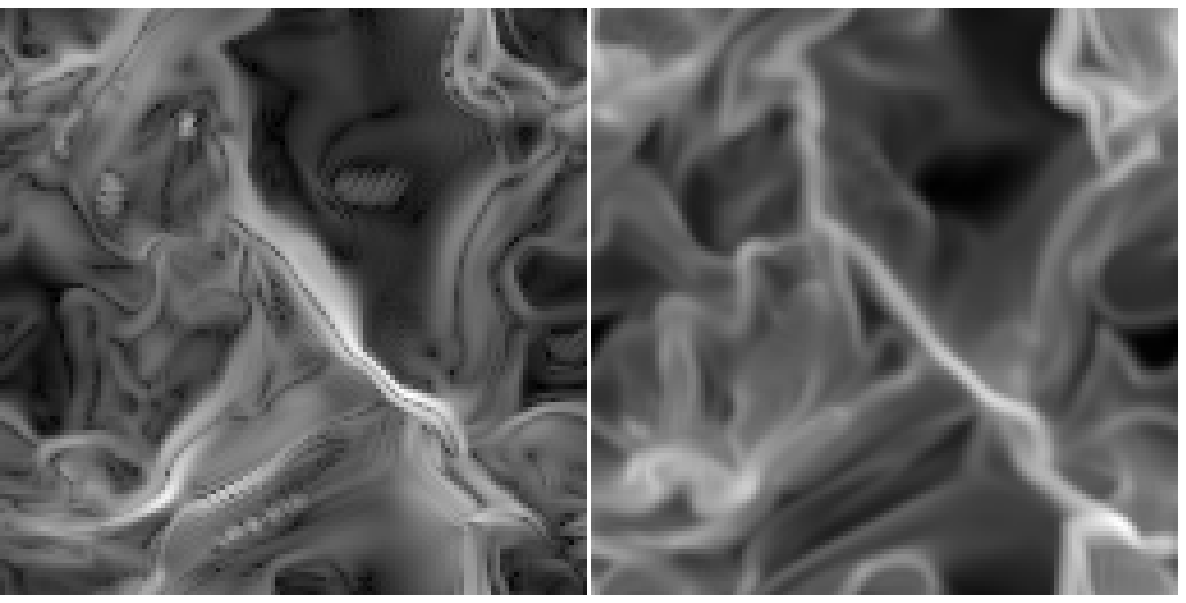}}
\caption[]{}
\label{fig7}
\end{figure}

{\bf Figure \ref{fig7}:}  Images of two--dimensional slices of ambipolar 
diffusion heating rate per unit volume (left) and of gas density (right),
from a snapshot of a 128$^3$ simulation, with ${\cal M}_A\approx 5$, and
$\langle |B|\rangle \approx 6\mu$~G. The intensity
in the image is proportional to $H_{AD}^{0.1}$ (left) and $n^{0.1}$
(right).  \\

\begin{figure}
\centerline{\epsfxsize=15cm \epsfbox{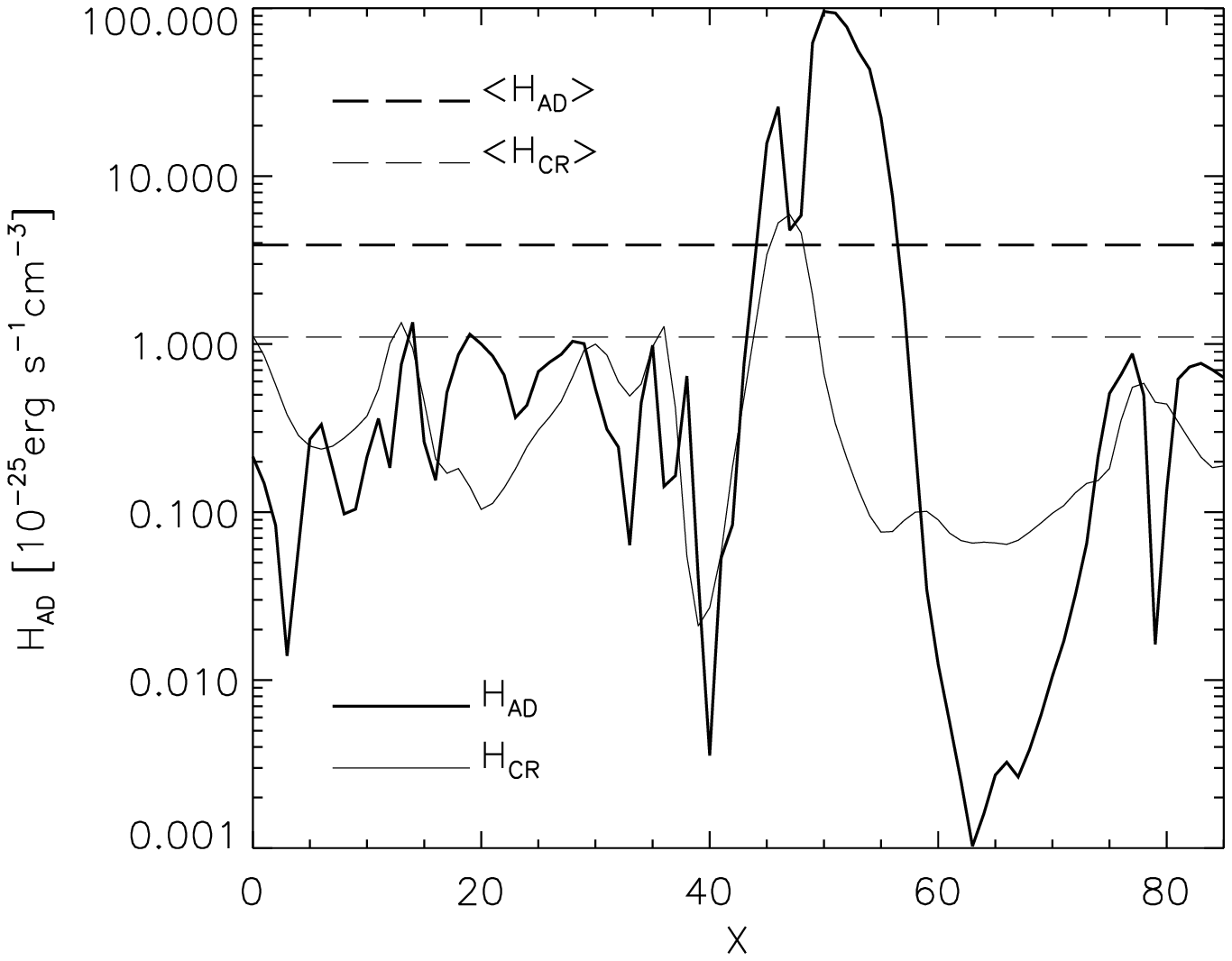}}
\caption[]{}
\label{fig8}
\end{figure}

{\bf Figure \ref{fig8}:} Local ambipolar drift heating rate
(thick line) and cosmic ray heating rate (thin line) per unit
volume. It is a cut through the slice presented in Fig.~\ref{fig7}.
The dashed lines show the values of the heating rates averaged over 
the whole computational box.\\

\begin{figure}
\centerline{\epsfxsize=15cm \epsfbox{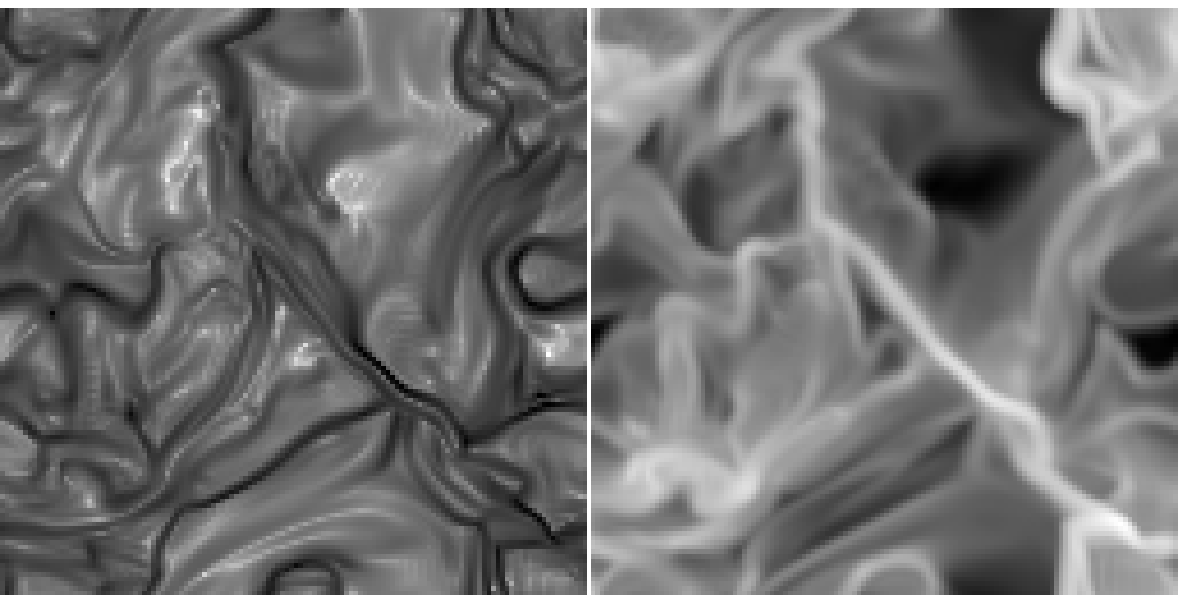}}
\caption[]{}
\label{fig9}
\end{figure}

{\bf Figure \ref{fig9}:}  Images of two--dimensional slices of magnetic length 
scale (left) and gas density (right). The slice is the same used in Fig.~\ref{fig7},
and the intensity in the image is proportional to $L_B^{0.1}$ (left) and $n^{0.1}$
(right).  \\

\clearpage
\begin{table}
\begin{tabular}{lrcrcrl}
\hline
\hline
 Model name  & ${\cal M}_A$ & $<|B|>/\mu$G & ${\cal M}$  &  $a$     & Size  &   Remarks  \\ 
\hline
         A1 ..............  &   83.1      &   0.4         &   9.9       &   0.3     & 128$^3$ &   Driven   \\ 
         A2 ..............  &   18.2      &   2.1         &  12.0       &   0.3     & 128$^3$ &   Driven  \\ 
         A3 ..............  &    8.2      &   3.5         &   9.1       &   0.3     & 128$^3$ &   Driven  \\ 
         A4 ..............  &    5.5      &   6.1         &  10.5       &   0.3     & 128$^3$ &   Driven  \\ 
         A5 ..............  &    4.2      &   8.1         &  10.8       &   0.3     & 128$^3$ &   Driven  \\ 
         A6 ..............  &    2.5      &  10.1         &   8.1       &   0.3     & 128$^3$ &   Driven  \\ 
         A7 ..............  &    0.7      &  59.5         &  12.4       &   0.3     & 128$^3$ &   Driven  \\
  A4$_{16}$ ...........  &    5.3      &   5.1         &   8.5       &   0.0375  &  16$^3$ &   Driven  \\ 
  A4$_{32}$ ...........  &    4.0      &   6.3         &   8.0       &   0.075   &  32$^3$ &   Driven  \\  
  A4$_{64}$ ...........  &    5.0      &   5.9         &   9.5       &   0.15    &  64$^3$ &   Driven  \\  
         B1 ..............  &    1.0      &  27.6         &   9.2       &   0.3     & 128$^3$ &   Decaying \\ 
 A3$_{.15}$ ...........  &    8.8      &   3.1         &   8.8       &   0.15    & 128$^3$ &   Driven  \\ 
 A3$_{.6}$ ............  &   10.6      &   3.1         &  10.5       &   0.6     & 128$^3$ &   Driven  \\ 
 A3$_{1.0}$ ...........  &    9.6      &   3.1         &   9.5       &   1.0     & 128$^3$ &   Driven  \\
\hline
\end{tabular}
\caption{}
\end{table}

\clearpage

\clearpage

\clearpage

\clearpage

\clearpage

\clearpage

\clearpage

\clearpage

\clearpage

\end{document}